# Commutator expansions in NMR diffusometry

Matias Nordin[a]

a) Medical Physics in Radiology, German Cancer Research Center, Heidelberg, Germany.

m.nordin@dkfz.de, +49 6221 42 2603

## Abstract

In this paper commutator expansions for solving the Bloch-Torrey's equation are derived. An exact solution for free diffusion in a constant magnetic field gradient is found. Furthermore the moments of the signal in the short gradient pulse limit for spins confined to a restricted geometry are derived. The moments are tested numerically against the known solution of diffusing spins between two perpendicular plates in the short gradient pulse limit.

1. Introduction

In 1966 Robertson presented an interesting solution to the problem of a diffusing spin confined to a bounded region in presence of a magnetic field gradient [1]. The problem was studied to gain understanding of the signal response when diffusion is measured using NMR in restricted media. In subsequent years, a wide range of solutions and approaches were developed to tackle the problem of confined diffusion in order to analyze the signal response in porous media [2-7]. Today NMR diffusometry has emerged to become a standard tool for characterizing diffusion in porous media and it is said that the motion of the diffusing particles indirectly probes the pore space geometry [8]. This has been particularly fueled by the pulsed gradient spin echo technique, where even diffractive effects have been observed in confined media [9]. In the pulsed gradient spin echo experiment, the experimentally obtained signal is plotted against $q$, a parameter that includes both the gradient strength as well as the gradient time. The initial slope of the logarithm of the signal ($q<<1$) is proportional to the diffusion coefficient of the diffusive species. Therefore, it is expected that by using a short experimental time, the slope of the logarithm of the signal is proportional to the free diffusion constant of the diffusive species. A deviance may be observed in porous media by considering longer experimental times, giving the diffusive species time to probe the confined geometry [10]. In an attempt to describe such deviance from the free diffusion constant, a power expansion of the signal has been considered. Using this power expansion, the expansion of the logarithm of the signal can also be formed, which typically is referred to as the cumulant expansion. An interesting feature of the cumulant expansion is that the fourth and higher order moments are all zero when the diffusive species undergo Gaussian diffusion and the second moment is proportional to the diffusion coefficient of the diffusive species. An estimate of the fourth (and higher) moments can therefore be made as an estimate of the deviance from Gaussian diffusion, for a given experimental time.

It is clear that the deviance from Gaussian diffusion gives additional information about the porous media, beyond the diffusion coefficient, in particular for longer diffusional times. Although it is not clear what information residues in this deviance, it is commonly thought to be related to *microstructural heterogeneity*. However, since a direct investigation of the relationship between the pore space geometry and the moments of the signal is challenging at best and a general theoretical framework connecting the pore space geometry and higher order moments is missing, the relationship between the deviance and microstructure remains strictly conceptual.

An interesting contribution in the context of developing a theoretical framework was presented in the nineties by Barzykin where the time integrated form of the Bloch-Torrey equation was investigated [3,11]. This approach was an extension of the work by Robertson and showed promise for analyzing the geometrical impact on the signal, since the theory was developed using an arbitrary geometry. Barzykin realized that the time integrated Bloch-Torrey equation quickly converges with respect to the number of eigenvalue/eigenfunction pairs used. Hence a good approximation to the problem may be reached using a finite number of eigenvalues/eigenvectors. This gave an important generalization, since the diffusion operator then does not need to be fully diagonalized for calculation of the magnetization and hence complex geometries could be investigated using numerically obtained eigenfunctions/eigenvalues [12,15]. The calculation then leads to a resulting matrix exponential that can numerically be solved for most problems on a standard laptop [11,12]. It is however still rather unsatisfactory that the neat analytical structure revealed using this approach is obscured in the numerical machinery of performing a matrix exponential. This makes it still difficult to analyze the resulting magnetization in terms of the pore space geometry impact.

In this paper, a novel approach is derived using commutator relations that extend the analytical structure found by Robertson and Barzykin to investigate NMR diffusometry in porous media. The approach is demonstrated by investigating free diffusion under a steady gradient as well as the calculation of the moments appearing in the cumulative expansion, using the short gradient pulse approximation.

2. **Theory**

The magnetization $m(\mathbf{r}, t)$ of a diffusing spin in presence of a magnetic field gradient can be described by the following Bloch-Torrey equation [13]

$$\dot{m}(\mathbf{r}, t) = (D_0 \Delta + i\gamma B(\mathbf{r}))m \quad r \in \Omega \qquad 1$$

$$\left(\frac{\partial}{\partial n} + b\right) m(\mathbf{r}, t) = 0 \qquad r \in \Gamma$$

Where $D_0$ denotes the self-diffusion constant of the spin, $\gamma$ the gyromagnetic ratio, $b$ the surface relaxivity and $B$ a function that describes the gradient term in space. In most applications the gradient term is linear in space and we may define the x-direction being the direction of the gradient, giving $B = gx$ where the scalar $g$ denotes the gradient strength. We will consider the case of perfect reflection and let therefore $b \to 0$. We will also, for ease of readability and notation, set $D_0 = \gamma = 1$ and note that the mathematical analysis remains the same. Eq. 1 is difficult to solve, due to both the fact that the operator described on the first line is not Hermitian and the given boundary conditions in the second line.

A commonly used approach to solve Eq. 1 is to begin by formally integrating it in time [3]. This yield

$$m(r, t) = e^{-t(\Delta_* + igx)} m(r, t = 0) \qquad 2$$

for some operator $\Delta_*$ that satisfies the boundary conditions in Eq. 1. In absence of boundary conditions, the operator equals the ordinary Laplace operator $\Delta_* \to \Delta$. If boundary conditions are present, the operator may conveniently be represented by its spectral expansion

$$\Delta_* = \sum_{n=0}^{\infty} |n\rangle \lambda_n \langle n| \qquad 3$$

Where $\lambda_n$ denotes an eigenvalue to $\Delta_*$ and $|n\rangle$ its corresponding eigenvector in standard bra-ket notation.

Given that the eigenvalues and eigenfunctions to $\Delta_*$ have been calculated (approximately or exactly) the solution to Eq. 2 can be calculated explicitly by the following representation as (infinite) matrices $A$ and $B$ in the following way

$$A_{nm} = D_0 \langle n | \Delta_* | m \rangle$$
$$B_{nm} = ig \langle n | x | m \rangle.$$

In this way Eq. 2 is expressed as

$$m(r,t) = e^{-t(A+B)} m_0$$

and the problem is reduced to a matrix operation (although the matrices are of infinite size) [3,11]. The matrix $A$ is diagonal and consists of the eigenvalues to $\Delta_*$ on the diagonal. The growing magnitude of the eigenvalues ensure that a good approximation to the solution of Eq. 2 can be found by truncating the two matrices $A$ and $B$ to a finite size $N \times N$ consisting of the integrals in Eq. 4 calculated for the eigenfunctions corresponding to the N first smallest eigenvalues of $\Delta_*$ [12]. Typically the last step in solving Eq. 5 is to numerically perform the matrix exponent for a given experimental time $t$.

However, it is rather unfortunate that the neat structure of $A$ and $B$ in calculating the resulting magnetization is hidden when performing the matrix exponent. This reduces the possibility of analyzing the effect on the magnetization by the geometry as well as gradient. It may be tempting to solve this issue by naively separating the exponent as $e^{A+B} = e^A e^B$, which would greatly simplify the analysis, but this is unfortunately not possible.

The reason for the exponent not to be separable is that the matrices (operators) $A$ and $B$ do not commute. The commutator is defined as $[A,B] = AB - BA$ and is another matrix (operator). If the two matrices (operators) do not commute, the exponent can still be expanded into the following series called the Zassenhaus formula [14]

$$e^{t(A+B)} = e^{tA} e^{tB} e^{-\frac{t^2}{2}[A,B]} e^{\frac{t^3}{6}(2[B,[A,B]] - [A,[A,B]])} \ldots$$

In our context, the higher order commutators can be viewed as a preparation of the initial magnetization, before it is subject to the operations of the exponentials $tB$ and finally $tA$. A similar expansion as the Zassenhaus formula may also be found in the calculation of the moments of the signal in a pulsed-gradient spin-echo experiment of restricted diffusion, when the short gradient pulse limit is applicable. The equation at hand is

$$E(q,t) = \langle q | e^{-t\Delta_*} | q \rangle$$

Where $E(q,t)$ denotes the echo decay and $e^{-t\Delta_*}$ denotes the diffusion propagator. The functions $|q\rangle$ denote the Fourier functions $|q\rangle = a_n \exp[iqx]$, where again the gradient is considered in the x-direction. The echo decay can be expanded in moments valid for small $q$

$$E(q,t) = \frac{E_0}{0!} + \frac{E_1}{1!} q + \frac{E_2}{2!} q^2 + \cdots$$

The moments are found by

$$E_n = \frac{\partial^n E(q,t)}{\partial q^n}$$

The derivative of the echo decay may be calculated from Eq. 7

$$\frac{\partial E(q,t)}{\partial q} = \frac{\partial}{\partial q}\langle q|e^{-t\Delta_*}|q\rangle = i2\pi\langle q|e^{-t\Delta_*}|q\rangle - iq\langle q|e^{-t\Delta_*}x|q\rangle \qquad 10$$

$$= i2\pi\langle q|[x, e^{-t\Delta_*}]|q\rangle$$

The second derivative gives

$$\frac{\partial^2 E(q,t)}{\partial q^2} = (i2\pi)^2 \langle q|[x,[x, e^{-t\Delta_*}]]|q\rangle \qquad 11$$

and so forth generating higher order commutators for increasing moments. In this case the diffusion is no longer free and the commutators must be evaluated using the spectral expansion of $\Delta_*$ in the following way. The commutators may be explicitly evaluated using expressions such as $\langle m|x|n\rangle = \alpha_{mn}$ forcing the operator $x$ to lie in the eigenspace of the diffusion propagator in order to ensure that the boundary conditions are satisfied [1]. The action of $x$ may in this way be represented as the matrix $\alpha_{mn}$, giving for the first commutator

$$\langle m|[x, e^{-t\Delta}]|n\rangle = \langle m|(xe^{-t\Delta_*}|n\rangle - e^{-t\Delta_*}x|n\rangle) =$$

$$(\langle m|e_0^{-t\lambda_n}X|n\rangle - e^{-t\Delta}X|n\rangle) = \qquad 12$$

$$(\langle m|(e^{-t\lambda_n}\sum_{k=0}^{\infty}|k\rangle\alpha_{kn} - \sum_{k=0}^{\infty}|k\rangle e^{-t\lambda_k}\alpha_{kn}))=$$

$$i2\pi\alpha_{mn}(e^{-t\lambda_n} - e^{-t\lambda_m}).$$

We may note that the commutator is time dependent, and that it actually is zero when $t = 0$. The second commutator is calculated in a similar way

$$\langle m|[X,[X, e^{-t\Delta}]]|n\rangle =$$

$$\langle m|XXe^{-t\Delta}|n\rangle - 2\langle m|Xe^{-t\Delta}X|n\rangle + \langle m|e^{-t\Delta}XX|n\rangle = \qquad 13$$

$$\beta_{mn}(e^{-t\lambda_n} + e^{-t\lambda_m}) - \sum_{k=0}^{\infty}\alpha_{mk}e^{-tk}\alpha_{kn}$$

and so forth for higher order moments. Note that the action of XX appearing in the second commutator is conveniently represented by a matrix $\beta_{mn}$. An explicit expression for the two first moments is

$$E_1 = i2\pi \sum_{n,m=0}^{\infty} \langle q|m\rangle \alpha_{mn}(e^{-t\lambda_n} - e^{-t\lambda_m})\langle n|q\rangle \qquad 14$$

$$E_2 = (i2\pi)^2 \sum_{n,m=0}^{\infty} \langle q|m\rangle \left(\beta_{mn}(e^{-t\lambda_n} + e^{-t\lambda_m}) - \sum_{k=0}^{\infty}\alpha_{mk}e^{-t\lambda_k}\alpha_{kn}\right)\langle n|q\rangle.$$

The expansion outlined in Eq. 8-14 is formal and possible to calculate when the spectrum of the diffusion operator is known, numerically or analytically. It may be noted that only the even moments contribute to the total signal in a NMR diffusometry experiment. The moments for the logarithm of the signal, can in a similar way be evaluated as a commutator expansion by combining the derivation in Eq. 8-14 with the Taylor expansion of $\ln|1 + x|$. One may also note that with perfect reflection at the boundaries only the first eigenfunction contributes in the limit of $q \to 0$ so the sum over $n$ and $m$ reduces to $m = n = 0$ with $\langle q|m\rangle = \langle q|n\rangle = 1$.

The use of these commutator expansions will now be demonstrated using two examples where the first example utilizes Zassenhaus formula for calculating free diffusion in presence of a constant gradient and a second example of non-free diffusion, where an expansion of the signal in the short gradient pulse limit is considered for the case of diffusing spins between two plates.

*Example 1: Free diffusion, constant gradient*

In the case of a constant gradient and absence of boundary conditions the operator $\Delta_*$ in Eq. 2 reduces to the ordinary Laplace operator and the equation at hand is

$$m(r,t) = e^{-t(\Delta + igx)} m(r, t = 0) \tag{15}$$

The two operators $\Delta$ and $x$ do not commute. We calculate the commutators and apply the Zassenhaus formula. The commutator between $\Delta$ and $x$ can be calculated by assuming an arbitrary test function $f(r)$

$$[\Delta, x] f(r) = \left( \frac{\partial^2}{\partial x^2} + \frac{\partial^2}{\partial y^2} + \frac{\partial^2}{\partial z^2} \right) x f(r) - x \left( \frac{\partial^2}{\partial x^2} + \frac{\partial^2}{\partial y^2} + \frac{\partial^2}{\partial z^2} \right) f(r) = \tag{16}$$

$$= \frac{\partial}{\partial x}(f + x f'_x) + x(f''_{yy} + f''_{zz}) - x \Delta f = 2 f'_x$$

$$\rightarrow [\Delta, x] = 2 \frac{\partial}{\partial x}$$

The higher order commutators yield

$$[\Delta, [\Delta, x]] f = \left[ \Delta, 2 \frac{\partial}{\partial x} \right] f = 2 \left( \Delta f'_x - \frac{\partial}{\partial x} \Delta \right) f = 0 \tag{17}$$

And

$$[x, [\Delta, x]] f = \left[ x, 2 \frac{\partial}{\partial x} \right] f = 2 \left( x f'_x - \frac{\partial}{\partial x}(xf) \right) = 2(xf' - f - xf') \tag{18}$$

$$\rightarrow [x, [\Delta, x]] f = -2I.$$

The identity operator is diagonal in all bases and hence commutes with all operators. Therefore the Zassenhaus formula gives a finite expression in terms of free diffusion. By combining Eq. 5 with Eq. 6 and using the above commutator relations, the following result is obtained

$$e^{-t(D_0 \Delta + igx)} = e^{-tD_0 \Delta} e^{-ti\gamma gx} e^{-\frac{t^2 D_0 i \gamma g \partial}{\partial x}} e^{-t^3 \frac{2}{3} D_0 \gamma^2 g^2}. \tag{19}$$

As we consider free diffusion, the domain is infinite and the Laplace operator has no spectrum. We are interested in the macroscopic signal and note that this is very simple in Fourier space, by the following observation [4]

$$m(t) = \int_\Omega m(r,t) = \lim_{q \to 0} \langle q | m(r,t) \rangle \tag{20}$$

By assuming that the initial distribution is flat in space ($|m_0\rangle = |q=0\rangle$) and combining Eq. 19 and 20 we obtain

$$m(t) \propto e^{-t^3 \frac{2}{3} D_0 \gamma^2 g^2} \lim_{q \to 0} \langle q | e^{-tD_0 \Delta} e^{-ti\gamma gx} e^{-t^2 D_0 i \gamma g \frac{\partial}{\partial x}} | q = 0 \rangle \tag{21}$$

By formally expanding the derivative term one realizes that the only contribution comes from the zeroth term, when the initial distribution is homogeneous and furthermore, the first order gradient term is easily closed in Fourier space, giving

$$\ldots = e^{-t^3 \frac{2}{3} D_0 \gamma^2 g^2} \int_{-\infty}^{\infty} \delta(k - gt\gamma) dk = e^{-t^3 \frac{2}{3} D_0 \gamma^2 g^2} \qquad 22$$

which is in accordance with previous results for a freely diffusing spin under subject of a linear magnetic gradient term (see e.g. [16,17]).

*Example 2: Diffusion between two plates, SGP-limit*

In the case of diffusion between two plates separated by a distance $a$, the eigenfunctions and eigenvalues of the diffusion propagator are analytically known [1]. For this example it is convenient to refine our dimensions as $t \to \frac{a^2}{D}$ and define the eigenfunctions on the range $x \in [0,1]$ as $|n\rangle = (2 - \delta_{n0})^{1/2} \cos(n\pi x)$. The corresponding eigenvalues become $\lambda_n = n^2 \pi^2$, for $n = 0,1,2,\ldots$ The calculation of the matrices $\alpha_{mn}$ and $\beta_{mn}$ in 14 may in fact be evaluated exactly. The expressions are

$$\langle m|x|n\rangle = i2\pi \alpha_{mn} = i2\pi \times \begin{cases} \dfrac{((-1)^{n+m} - 1)(m^2 + n^2)\sqrt{2 - \delta_{0m}}\sqrt{2 - \delta_{0n}}}{\pi^2 (m^2 - n^2)^2}, & n \neq m \\ \dfrac{1}{4}(2 - \delta_{0m}), & n = m \end{cases} \qquad 23$$

and

$$\langle m|xx|n\rangle = (i2\pi)^2 \beta_{mn} = \qquad 24$$

$$(i2\pi)^2 \times \begin{cases} \dfrac{2(-1)^{n+m}(m^2 + n^2)\sqrt{2 - \delta_{0m}}\sqrt{2 - \delta_{0n}}}{\pi^2 (m^2 - n^2)^2}, & n \neq m \\ \dfrac{1}{3} + \dfrac{\frac{1}{2}((-1)^{\delta_{0n}} + 1)}{2\pi^2 n^2}, & n = m \end{cases}$$

One may note that the sums in Eq. 14 quickly converge for $t > 0$ due to the increased magnitude of the eigenvalues. It is instructive to investigate the $\langle m|Xe^{-t\Delta}X|n\rangle$ term in the second moment, as it traces over the whole diffusion kernel, why the elements in the sum over the $k$-index in Eq. 14 are denoted as

$$M(k) = \alpha_{mk} e^{-t\lambda_k} \alpha_{kn}. \qquad 25$$

In practice it is not necessary to include many terms when $t \gg 1$ and the residual may easily be checked by using the exponent $e^{-t\lambda_k}$ for the maximum $k$ used.

## 3. Results and discussion

In Fig. 1 the elements $M(k)$ in Eq. 25 for the case of diffusion between two plates are plotted for $t = 0$ as well as for an increased time and show the finiteness of the trace as $t > 0$. In Fig. 2, the approximate signal formed by choosing a finite number of moments (Eq. 8) is plotted against $qa$ up to $qa = 0.5$ and a convergence towards the analytic echo decay when the number of moments used increase. One may note that the signal deviates substantially from the analytic solution at $\frac{ta^2}{D} = 1$ when only the second moment is present, and in fact also still does so when the first four moments are included. From the six first moments however, the signal starts to converge below $< 0.5$. The derived expressions for the moments do not rely explicitly on the knowledge of the analytic form of the eigenfunctions/eigenvalues of the underlying diffusion operator, so numerical calculations of such could be considered in more complex situations. The resulting moments are expressed as a sum over the diffusion propagator and are straight forward to evaluate given that the eigenvalues and eigenfunctions of the diffusion operator are known.

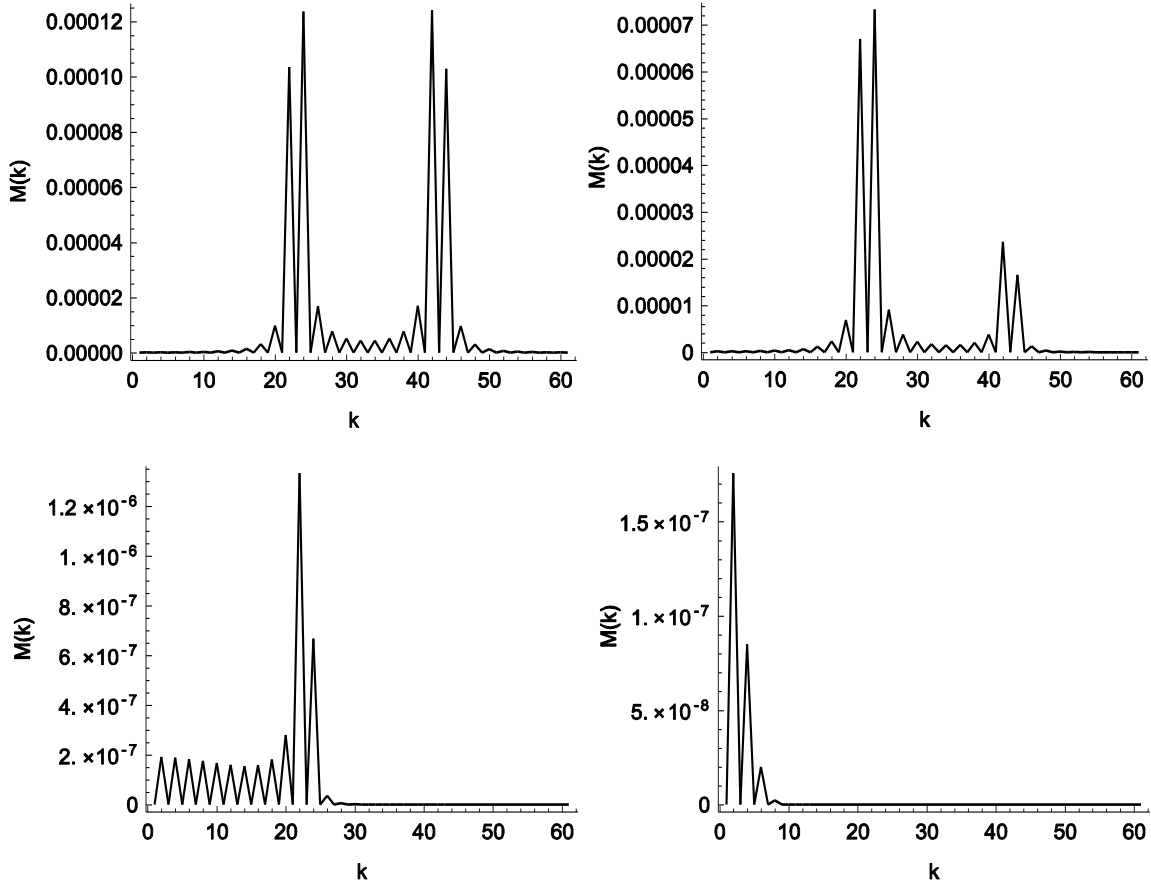

Figure 1. The sum over the diffusion kernel in calculating the second moment quickly decay as k grows. The figure show the elements of the sum $M(k) = \alpha_{mk} e^{-t\lambda_k} \alpha_{kn}$ (Eq. 25) calculated for diffusion between two plates for $t = 0$ (top left), $t = 1 \times 10^{-4}$ (top right), $t = 1 \times 10^{-3}$ (bottom left) and $t = 1 \times 10^{-2}$ (bottom right) for $m = 22$ and $n = 42$.

The exponential convergence with respect to the number of eigenvalues/eigenfunctions used ensures that a good approximation may be found by a finite set of eigenvalues/eigenfunctions. Hence it is not necessary to fully diagonalize the diffusion operator when moments are calculated in the short gradient pulse limit in restricted media. The approach of calculating moments is not necessary dependent on a specific geometry where the eigenvalues/eigenfunctions are analytically known, but more complex geometries could be considered using numerical solutions to the smallest eigenfunctions/eigenvalues in combination with the procedure outlined in Eq. 10-14.

We may finally note that the calculation of the sum with index $k$ in Eq. 14 as $t \to 0$ may be solved analytically for geometries where the eigenfunctions of the diffusion operator are known using regularization techniques. Such type of calculation has been discussed previously by Grebenkov (see appendix in [10]).

## 4. Conclusions

In this study, the Bloch-Torrey equation is investigated using commutator relations and a method for deriving moments in the short gradient pulse limit is presented. The approach relies on the spectrum of the diffusion operator and is therefore applicable to any geometry where the Laplace eigenvalues and eigenfunctions can be calculated, analytically or numerically.

## 5. Acknowledgements

The author would like to thank the support of the organizers of ISMRM12 in the form of a travel grant, which enabled the attendance of the conference and Dr. Frederik Laun and Dr. Bram Stieltjes for providing with valuable comments.

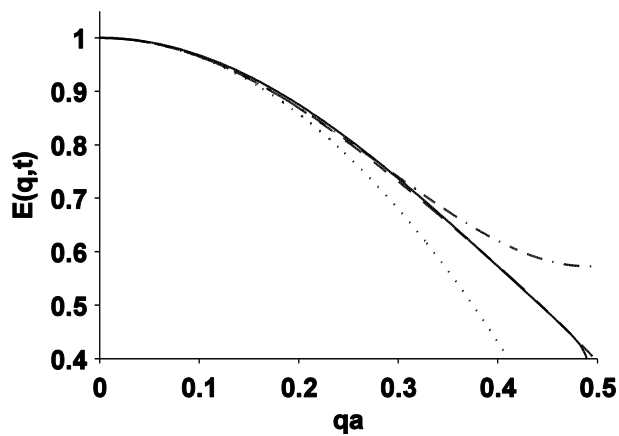

Figure 2. Truncated moment expansions of the signal calculated using Eq. 9-12 for the problem of diffusing spins between two perpendicular planes separated by $a$ in the SGP-limit. The first two moments (dotted), the first four moments (dashed dotted) deviate from the analytical signal (fully drawn line) at higher $qa$-values, while the first six moments (dashed) almost coincide with the analytical solution. The signals are calculated for $Dt/a^2 = 1$.

**6. References**

[1] B. Robertson, Phys. Rev. **151** (1966) 273-277

[2] C. H. Neuman, J. Chem. Phys. **60** (1974) 4508-4511

[3] A. V. Barzykin, Phys. Rev. B **58** (1998) 14171

[4] V. M. Kenkre, E. Fukushima and D. Sheltraw, J. Magn. Reson. **128** (1997) 62–69

[5] P. T. Callaghan, J Magn Reson. **129:1** (1997) 74-84.

[6] A. F. Frøhlich, L. Østergaard and V. G. Kiselev, J. Magn. Reson. **179:2** (2006) 223-33

[7] P. Linse and O. Söderman, J. Magn. Reson. A **116** (1995) 77

[8] P. P. Mitra, N. S. Pabitra, L. M. Schwartz and P. Le Doussal, Phys. Rev. Lett. 68 (1992) 3555

[9] P. T. Callaghan, A. Coy, D. Macgowan, K. J. Packer and F. O. Zelaya, Nature 351 (1991) 467-469

[10] D. S. Grebenkov, Rev. Mod. Phys. **79** (2007) 1077

[11] A.V. Barzykin J. Magn. Reson. **139** (1999) 342

[12] M. Nordin, D. S. Grevenkov, M. Nilsson-Jacobi, and M. Nydén, Microporous Mesoporous Mater. 178 (2013) 7–10

[13] H. C. Torrey, Physical Review **104:3** (1956) 563

[14] W. Magnus, Comm. Pure Appl. Math **7:4** (1954) 649–673

[15] D. S. Grebenkov, Diffusion Fundamentals **5:1** (2007) 1-34

[16] E. Hahn. Physical review **80:4** (1950) 580-594

[17] W.S. Price NMR studies of translational motion, Cambridge University press, 2009